\documentclass[conference]{IEEEtran}

\ifCLASSINFOpdf
\else
\fi

\ifCLASSINFOpdf   
   \usepackage[pdftex]{graphicx} 
   \graphicspath{{../pdf/}{../jpeg/}{./image/}}    
   \DeclareGraphicsExtensions{.pdf,.jpeg,.png,.jpg}   
 \else   
   \usepackage[dvips]{graphicx}   
   \graphicspath{{../eps/}}    
\fi
\usepackage{float}

\usepackage{amsmath}

\usepackage{algorithm}

\usepackage{algpseudocode}
\begin{document}
\setlength{\parskip}{0pt} 
\title{Full-Duplex Transceiver for Future Cellular Network: A Smart Antenna Approach\vspace{-1.0em}}
\author{\IEEEauthorblockN{Chandan Pradhan and Garimella Rama Murthy}
\IEEEauthorblockA{Signal Processing and Communication Research Center\\
International Institute of Information Technology, Hyderabad, India \\
Email: chandan.pradhan@research.iiit.ac.in, rammurthy@iiit.ac.in}
}


%


\maketitle

\begin{abstract}
In this paper, we propose a transceiver architecture for full-duplex (FD) eNodeB (eNB) and FD user equipment (UE) transceiver. For FD communication,.i.e., simultaneous in-band uplink and downlink operation, same subcarriers can be allocated to UE in both uplink and downlink. Hence, contrary to traditional LTE, we propose using single-carrier frequency division multiple accesses (SC-FDMA) for downlink along with the conventional method of using it for uplink. The use of multiple antennas at eNB and singular value decomposition (SVD) in the downlink allows multiple users (MU) to operate on the same set of subcarriers. In the uplink, successive interference cancellation with optimal ordering (SSIC-OO) algorithm is used to decouple signals of UEs operating in the same set of subcarriers. A smart antenna approach is adopted which prevents interference, in downlink of a UE, from uplink signals of other UEs sharing same subcarriers. The approach includes using multiple antennas at UEs to form directed beams towards eNode and nulls towards other UEs. The proposed architecture results in significant improvement of the overall spectrum efficiency per cell of the cellular network.\par
\end{abstract}


%
\IEEEpeerreviewmaketitle

\section{Introduction}

One of the major revolutions in the future wireless networks can be the introduction of full-duplex (FD) eNodeB (eNB) and FD user equipment (UE). In traditional duplexing, either two separate channels or time slots are used for uplink and downlink. FD systems make the simultaneous in-band transceiving feasible. In recent years, excessive work is being done in the area of self-interference cancellation (SIC) design for both single and multiple antenna transceiver units \cite{full_duplex, FD_small}. This enables optimal cancellation of interference from the receiver chains introduced by the transmitter chains of the transceiver unit. Here, we discuss the FD transceiver design for eNB and UE and the corresponding uplink and downlink operations. \textit{We assume perfect self-interference cancellation at eNB and UE transceiver circuits.} While this is far from true today, sufficient progress is being made in this direction to start considering this model and its implications ,especially in case of  small cells, where the transmission power varies from 17dBm to 30dBm. In \cite{full_duplex}, the SIC design is capable of canceling the self-interference almost to the noise floor for a multiple antenna Wifi with transmission power ranging from 16dBm to 20dBm and bandwidth of 20MHz. This is a positive indicator for future of FD in cellular networks.\par
     In conventional LTE system, UEs are allocated sub carrier resources according to channel state scheduling algorithm \cite{sc_fdma}. For uplink and downlink, single carrier frequency division multiple access (SC-FDMA) and orthogonal frequency division multiple access (OFDMA) is used for multiple access respectively. For FD operation, same subcarriers can be allocated to UEs for uplink and downlink. Hence we propose using SC-FDMA for both uplink and downlink, due to its advantage, over OFDMA, in terms of bit error rate (BER) performance and energy efficiency, particularly at the UE. \cite{sc_fdma}. \par
    The FD operation enables channel reciprocity for uplink and downlink. This eases the availability of channel state information (CSI) at the transmitter. This, coupled with the use of multiple antennas at eNB allows multiple user (MU) to operate on the same set of subcarriers. The use of the same subcarriers for both uplink and downlink results in interference in downlink of a UE from uplink signals of other UEs operating in the same subcarriers. In our previous work \cite{WTS}, we neglected this interference assuming a non-line of sight (NLOS) scenario between UEs. For this, we considered an environment where a plethora of man-made and natural obstructions are present, like buildings and trees, between the UEs. This leads to screening of signals between the UEs. However, in this work, we consider a scenario for small cells, deployed around the lower end of the super high frequency (SHF) band (2 GHz to 7 GHz \cite{mimo_ofdm}), with possible LOS between the UEs sharing the same subcarriers. Hence a smart antenna based approach is deployed which uses multiple antennas at UEs \footnote{Operation in the GHz range considered, ease the deployment of multiple antennas at UE. Also extensive research is under way in the area of mmWaves \cite{mm_wave} which further facilitates deployment of multiple antennas at UE. Small cell deploying mmWaves constitutes the next part of the work.}  to form directed beams towards eNB and nulls toward other UEs coexisting in the same subcarriers \cite{beamforming}.  \par
    The paper is divided into six sections. In section 2, we will discuss about the system model describing the eNB and the UE design for the proposed method. Section 3 and section 4 deal with the downlink and uplink operations respectively. Simulation results for downlink and uplink are discussed in section 5. The conclusion is presented in Section 6. \par
    Notation: $[.]^T$,$(.)^H$ denote transpose and Hermitian respectively. $||.||_2$  denotes Euclidean norm. \par
    
\section{SYSTEM MODEL}

For facilitating FD communication, both eNB and UE operate in FD mode. The SIC cancellation design deployed at the RF front end of both eNB and UEs will allow simultaneous allotment of a single channel for both uplink and downlink operations \cite{full_duplex}. The self-interference at a receive chain of a multiple antenna system consists of 1)“self-talk”, .i.e., interference from the transmit chain with which receive chain shares antenna and 2) “cross-talk”, .i.e., interference from neighboring transmit chains. There are three major components for self-talk and cross talk \cite{full_duplex}: 1) Linear component, 2) Non-linear component (resulting when the digital baseband signal is converted to analog and up-converted to the carrier frequency by various analog circuits involved) and 3) Transmit Noise \cite{full_duplex}. The SIC design cancels out these components from the receive chains. The details of SIC design are not discussed here. In this work, we consider a \textit{single cell} deploying an eNB with $N_e$ antennas and $K$ UEs with $N_r$ antennas operating in FD mode. For the proposed transceiver architecture shown in  Fig.1 and Fig.2 for eNB and UE respectively, the Analog and Digital SIC unit, includes analog and digital cancellation stages described in \cite{full_duplex} for implementing SIC.  \par    
 \begin{figure}
\centering
\includegraphics[width=3.25in ,height=2.0in]{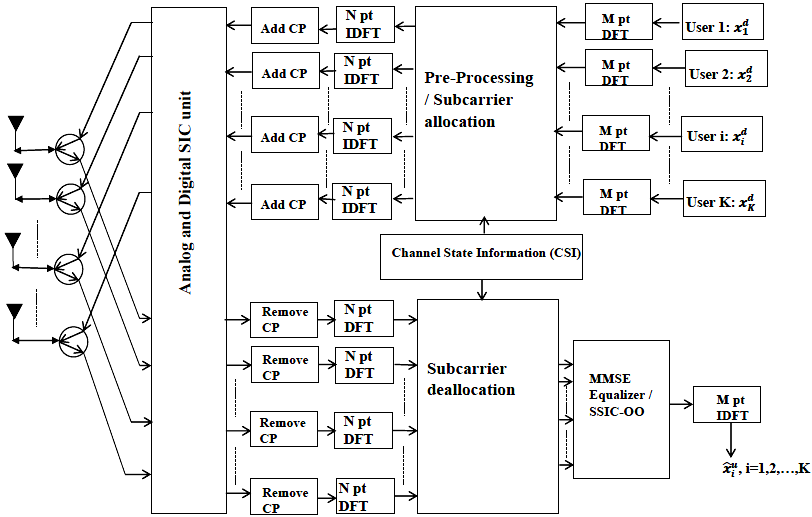}
\vspace{-0.5em}
   \caption{Transceiver structure for the proposed eNB architecture}
   \label{fone}
\end{figure}

\begin{figure}
\centering
\includegraphics[width=3.25in ,height=1.5in]{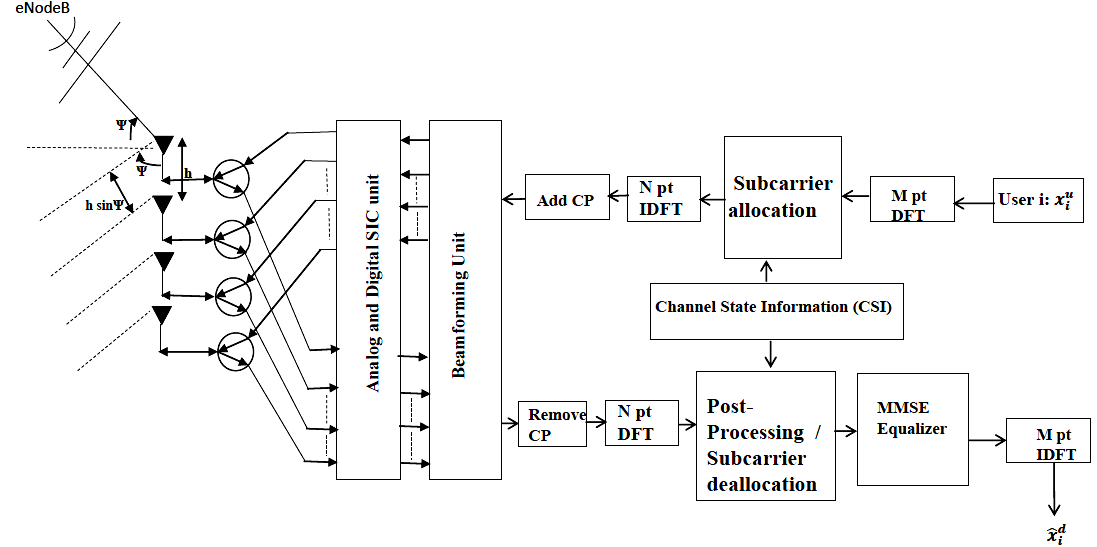}
\vspace{-0.5em}
   \caption{Transceiver structure for the proposed UE architecture}
   \label{ftwo}
   \vspace{-1.5em}
\end{figure}

    We assume dynamic subcarrier allocation based on channel state information. Let each subcarrier allocated be shared by $\acute{K}$ UEs simultaneously, where $\acute{K}$ is given by \cite{sc_fdma}     
\begin{equation}
\label{one}
 \acute{K} = \begin{cases} min\Big( \left \lfloor{\frac{N_e}{N_r}}\right \rfloor ,K \Big), N_e > N_r\\ 
                           1, otherwise \\ 
                           \end{cases}
       \end{equation}

\textit{ Note: In this work the $N_r$ antenna elements in the UE, unlike in case of eNB, are taken closely spaced enough to allow a high spatial correlation between them \footnote{For a small cell deployment, in the band under consideration, angular spread can be around $20^o$ for outdoor and $22^o-26^o$ for indoors\cite{mimo_ofdm}. Thus, by keeping antenna spacing below half wavelength, results in high correlation among antenna elements \cite{correlation}.}. Hence, UE acts as a single antenna system, i.e. $N_r \approx 1$, when evaluating $\acute{K}$. This eliminates the possibility of diversity gain at the UE.}\par
     Each UE is allocated M (=$\left \lfloor{\frac{N\acute{K}}{K}}\right \rfloor$ ) subcarriers, where N is the total number of subcarriers available.  Keeping this in mind, we here consider a case of $\acute{K}=K$, i.e., all the $K$ UEs are allocated all the $N$ subcarriers. The channel between each eNB antenna and UEs antenna is assumed to be frequency selective with $L$ taps. The FD operation allows the channel reciprocity between downlink and uplink:
                \begin{equation}
\label{two}
h^{u}_{j,i,k} (b) = h^{d}_{j,i,k} (b)
       \end{equation}
where  $ h^{u}_{j,i,k} (b)$ and $h^{d}_{j,i,k} (b) $  denotes $b^{th}$ time domain  uplink and downlink channel coefficient between $j^{th}$  antenna at eNB and $k^{th}$  antenna of $i^{th}$ UE respectively, $b = 0, 1, 2, ... , L-1$, $j = 1 , 2, ..., N_e$, $k = 1, 2,...,N_r $ and $i = 1,2, ...,K$. \par
     In downlink, the channel reciprocity property of FD enables the transmitter (eNB) to acquire CSI with ease. The CSI can be used to perform efficient subcarrier allocation and precoding the UE data at eNB so as to perform SVD based beamforming.\textit{ For FD operation, same subcarrier is allocated for both downlink and uplink.} In the uplink, successive interference cancellation with optimal ordering (SSIC-OO) algorithm is used at the eNB to segregate signals of UEs sharing the same subcarriers. Multiple antennas at UE are exploited to avoid interference to the UE in downlink from uplink signals of other UEs, sharing same subcarriers \footnote{ In a multicell scenario, intercell interference (eNB to eNB and eNB to UE) can be mitigated by methods like interference management through cloud access network (C-RAN) architecture. Discussion on inter cell interference is out of scope of this paper.}. The highly correlated multiple antennas UE are used to form a directed beam towards the eNB and nulls in the direction of other UEs operating in the same subcarriers. \textit{The directed beam also helps in combating high pathloss}\cite{mm_wave}. This is implemented through the beamforming unit shown in Fig.3.  To keep the analysis simple, the frequency domain MMSE equalizer is used both in downlink and uplink.\par

\begin{figure}
\centering
\includegraphics[width=2.5in ,height=1.5in]{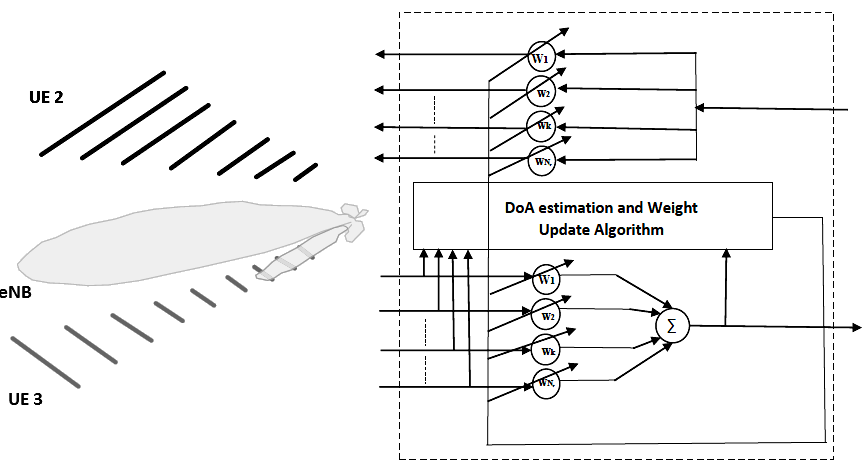}
\vspace{-0.5em}
   \caption{Structure for the Beamforming unit for the proposed UE architecture and the formation of directed beam toward eNB}
   \label{fthree}
   \vspace{-1.0em}
\end{figure}

\section{Downlink Operation}
    In Fig.\ref{fone}, let $\mathbf{x}^i_d$  denote the $i^{th}$ UE information data block of length $M$ in downlink (denoted by d):  \vspace{-1.0em}
    
\begin{equation}
\label{three}
\mathbf{x}_i^d = [x_i^d(1),x_i^d(2),...,x_i^d(M)]^T, i=1,2,...,K
       \end{equation}           
          
The output of the M-point block is given by: \vspace{-1.0em}

\begin{equation}
\label{four}
\mathbf{\bar{x}}_i^d = \mathbf{F}_M \mathbf{x}_i^d
       \end{equation}      
                                                                       
where $\mathbf{F}_M$  is the M-point DFT matrix and $ \mathbf{\bar{x}}_i^d = [\bar{x}_i^d(1),\bar{x}_i^d(2),...,\bar{x}_i^d(M)]^T$ . This output is then passed through the subcarrier allocation block. Let  $\mathbf{P}^i_m$ denotes $N_e X 1$  precoding vector for $i^{th}$ UE on $m^{th}$ subcarrier. The precoded output vector of size $N_e X 1$ for $i^{th}$ UE on $m^{th}$ subcarrier is given by: \vspace{-1.0em}

\begin{equation}
\label{five}
\mathbf{z}^d_i = \mathbf{P}^i_m \bar{x}_i^d (m)
       \end{equation}   
       
where $m=1,2,...,M$, $\mathbf{z}^d_i(m) = [z^d_{i,1}(m), z^d_{i,2}(m),...,z^d_{i,N_e}(m)]^T$ .  Let $\mathbf{A}^i$, represents the $N X M$ subcarrier allocation matrix for $i^{th}$ UE, $i=1,2,...,K$ [3]. The $N X 1$ vector input to the N-point IDFT block for $j^{th}$ transmit chain is given by:
\vspace{-1.0em}
\begin{equation}
\label{six}
\mathbf{e}^d_j = \sum_{i=1}^{K} \mathbf{A}^i \mathbf{z}^d_{i,j}
\end{equation}   
                                                 
where $ j=1,2,...,N_e$, $\mathbf{z}^d_{i,j} = [z^d_{i,j} (1),z^d_{i,j} (2) ,..., z^d_{i,j} (M)]^T$  . The output of the $N$-point IDFT block for the $j^{th}$ transmit chain is given by: \vspace{-1.5em}
                                               
\begin{equation}
\label{seven}
\mathbf{s}^d_j = \mathbf{\bar{F}}_N \mathbf{e}^d_j
\vspace{-0.10em}
       \end{equation}   
                                                       
where $j=1,2,...,N_e$, $\mathbf{\bar{F}}_N$ denotes the N-point IDFT matrix. This signal is then transmitted on $j^{th}$ antenna after addition of the cyclic prefix (CP). \par

In this work, the UE transceiver unit (Fig.\ref{ftwo}) consists of a uniformly spaced linear antenna array of $N_r$ elements with an inter element distance of $h$. The angle with respect to the array normal at which the plane wave impinges upon the array is represented as $\psi$. Let $\mathbf{A}=[ \boldsymbol{\alpha}(\psi_1), \boldsymbol{\alpha}(\psi_2),...,\boldsymbol{\alpha}(\psi_j),...,\boldsymbol{\alpha}(\psi_K)]$ be the $N_r X K$ steering matrix such that $\boldsymbol{\alpha}(\psi_j)$ is the $N_r X 1$ vector which represents the array steering vector corresponding to the direction of arrival ($\psi_j$) of either the eNB  or one of the other $K-1$ UEs sharing the same subcarriers \cite{beamforming}:\vspace{-0.915em}

\begin{equation}
\label{eight}
\boldsymbol{\alpha}(\psi_j)=[\alpha^0_j, \alpha^1_j, ..., \alpha^x_j, ..., \alpha^{{N_r}-1}_j]^H
       \end{equation}   
       
where  $\alpha^x_j = \exp(-j2 \pi x \frac{h}{\lambda} \sin(\psi_j))$ and  $\lambda$ is the wavelength. Algorithms like Root-MUSIC (due less computational complexity) is used for estimating DoA of eNB and other UEs sharing the same spectrum resource. By knowing the direction of the eNB, antenna array forms directed beams in its direction with a constant gain and null towards other UEs. Directions of other UEs are calculated using the interfering uplink signals. \par
     At the $l^{th}$   UE, the downlink received signal is obtained after SIC cancellation in receive chain. The received signal at $k^{th}$ antenna, is given by:\vspace{-1.5em}
     
      \begin{multline}
\label{nine}
\mathbf{y}^d_{l,k}= \Big [ \sum_{j=1}^{N_e} \mathbf{h}^d_{j,l,k} \otimes \mathbf{s}^d_j + \mathbf{n}_{l,k} \Big ] \alpha^{k-1}_{e,l} \\
+ \Big [ \sum_{\substack{ q=1 \\ q \neq 1}}^{K} \sum_{\acute{k}=1}^{N_r} \mathbf{h}^u_{q,\acute{k},l,k} \otimes \mathbf{s}^u_{q,\acute{k}} + \mathbf{\acute{n}}_{q,l,k} \Big ] \alpha^{k-1}_{q,l} 
       \end{multline}   
       
where $ l=1,2,...,K$, $k=1,2,...,N_r$. $\otimes$  denotes circular convolution operation. $\mathbf{y}^d_{l,k}$ is $N_{cp}X1$ vector where $N_{cp}$ is the receive symbol size with CP. $\mathbf{h}^{d}_{j,l,k}$ is the channel coefficient between $j^{th}$ antenna of eNB and $k^{th}$ antenna of $l^{th}$ UE in downlink.  $\mathbf{n}_{l,k}$ is the noise introduced in the channel between eNB and $k^{th}$ antenna of $l^{th}$ UE. $\alpha^{k-1}_{e,l}$ is the spatial response of the $k^{th}$ antenna of $l^{th}$ UE in the DoA of $\psi_{e,l}$. $\psi_{e,l}$ is the DoA of eNB w.r.t $l^{th}$ UE. $\mathbf{h}^u_{q,\acute{k},l,k}$ is the channel coefficient between $\acute{k}^{th}$ antenna of $q^{th}$ UE and $k^{th}$ antenna of $l^{th}$ UE. $\mathbf{s}^u_{q,\acute{k}}$ is the uplink signal from the $\acute{k}^{th}$ antenna of $q^{th}$ UE. $\mathbf{\acute{n}}_{q,l,k}$ is the noise introduced in the channel between $q^{th}$ UE and $k^{th}$ antenna of $l^{th}$ UE.  $\alpha^{k-1}_{q,l}$ is the spatial response of the $k^{th}$ antenna of $l^{th}$ UE in the DoA of  $\psi_{q,l}$. $\psi_{q,l}$  is the DoA of $q^{th}$ UE w.r.t $l^{th}$ UE. \par

In this paper, for the purpose of forming directed beam towards eNB, we have used the Constrained Least Mean Square algorithm (CLMS) due to its computational simplicity. The CLMS algorithm for determining the optimal weight vector for the look direction is \cite{beamforming}:\vspace{-1.0em}

\begin{equation}
\label{ten}
\mathbf{W}_l(p)=\mathbf{W}_l(p-1)+2 \mu (\mathbf{\dot{y}}^d_l(p-1))^T \mathbf{y}^{d}_l(p-1)
       \end{equation}   
       
       \begin{equation}
       \label{eleven}
\mathbf{W}^H_l \boldsymbol{\alpha}(\psi_x) \approx \begin{cases} 1 , x = j \\ 
                           0, x \neq j\\ 
                           \end{cases}, \forall p
       \end{equation}   
          
where $\mathbf{W}_l(p)=[w_{l,1}(p),w_{l,2}(p),...,w_{l,N_r}(p)]^T$ is the complex weight vector for the $l^{th}$ UE, in the $p^{th}$ iteration, $\boldsymbol{\alpha}(\psi_j)$ is the array response on the desired look direction at $\psi_j$. $\mathbf{\dot{y}}^d_l(p-1)=[\mathbf{y}^d_{1,l}(p-1),\mathbf{y}^d_{l,2}(p-1),...,\mathbf{y}^d_{l,N_r}(p-1)]$ is $N_{cp} X N_r$ matrix and $\mathbf{y}^d_l (p-1)$ is the weighted sum of the signals from the $N_r$ antennas in the $(p-1)^{th}$ iteration. The ‘$\mu$’ is a positive scalar, called gradient step size that controls the convergence rate of the algorithm. Henceforth, for the sake of analysis, the notion for the number of iteration $(p)$ will be ignored. All the notations will be taken to be at the $p^{th}$ iteration, which is large enough for the CLMS algorithm to converge.   The weighted sum of the outputs of the $N_r$ antennas is then calculated as follows:\vspace{-1.5em}
            
             \begin{multline}
       \label{twelve}
       \mathbf{y}^d_l = \sum_{k=1}^{N_r} w_{l,k}\mathbf{y}^d_{l,k}  = \sum_{k=1}^{N_r} \Big [ \sum_{j=1}^{N_e} \mathbf{h}^d_{j,l,k} \otimes \mathbf{s}^d_j + \mathbf{n}_{l,k} \Big ] w_{l,k} \alpha^{k-1}_{e,l} \\
+ \sum_{k=1}^{N_r} \Big [ \sum_{\substack{ q=1 \\ q \neq 1}}^{K} \sum_{\acute{k}=1}^{N_r} \mathbf{h}^u_{q,\acute{k},l,k} \otimes \mathbf{s}^u_{q,\acute{k}} + \mathbf{\acute{n}}_{q,l,k} \Big ] w_{l,k} \alpha^{k-1}_{q,l} 
       \end{multline}
       
Due to the high correlation between antennas at the UEs, these can be approximated as a single antenna system and hence (\ref{twelve}) is equivalent to the following \footnote{Due to space limitation, the proof is not included here.}:    \vspace{-1.0em}

                      \begin{multline}
       \label{thirteen}
       \mathbf{y}^d_l = \sum_{j=1}^{N_e} \Big ( \mathbf{h}^d_{j,l} \otimes \mathbf{s}^d_j + \mathbf{n}_{l} \Big ) \Big [ \sum_{k=1}^{N_r} w_{l,k} \alpha^{k-1}_{e,l} \Big ] \\
+ \sum_{\substack{ q=1 \\ q \neq 1}}^{K} \sum_{\acute{k}=1}^{N_r}   \Big ( \mathbf{h}^u_{q, l} \otimes \mathbf{s}^u_{q} + \mathbf{\acute{n}}_{q,l} \Big )  \Big [\sum_{k=1}^{N_r}w_{l,k} \alpha^{k-1}_{q,l} \Big ] 
       \end{multline}
       
where, for the $l^{th}$UE, $\mathbf{h}^d_{j,l,1} \approx \mathbf{h}^d_{j,l,2} \approx ...\approx \mathbf{h}^d_{j,l,N_r}\approx \mathbf{h}^d_{j,l}$, and $\mathbf{n}_{l,1} \approx \mathbf{n}_{l,2}\approx ... \approx \mathbf{n}_{l,N_r} \approx \mathbf{n}_l $. For the $q^{th}$ UE,$\mathbf{h}^u_{q,1,l,k} \approx \mathbf{h}^u_{q,2,l,k} \approx ...\approx \mathbf{h}^u_{q,N_r,l,k} \approx \mathbf{h}^u_{q,l,k}$, $\mathbf{h}^d_{q,l,1} \approx \mathbf{h}^d_{q,l,2} \approx ... \approx \mathbf{h}^d_{q,l,N_r} \approx \mathbf{h}^d_{q,l}$, $s^u_{q,\acute{k}}=\alpha^{\acute{k}-1}_{e,q} w_{\acute{k},q} s^u_q $ and $\mathbf{\acute{n}}_{q,l,1} \approx \mathbf{\acute{n}}_{q,l,2} \approx...\approx \mathbf{\acute{n}}_{q,l,N_r} \approx\mathbf{\acute{n}}_{q,l} $. 
                
Also, from (\ref{eleven}), received signal in (\ref{thirteen}) after removing of CP can now be given by:    
 \vspace{-0.5em}                               
                                  \begin{equation}
       \label{fourteen}
       \mathbf{y}^d_l=\sum_{j=1}^{N_e}\mathbf{h}^d_{j,l} \otimes \mathbf{s}^d_j + \mathbf{n}_{l}
       \vspace{-0.50em}
       \end{equation}
       
where $l=1,2,...,K$, $\mathbf{y}^d_l$ is of size $N X 1$, $\mathbf{h}^d_{j,l}=[h^d_{j,l}(0),h^d_{j,l}(1),...,h^d_{j,l}(L-1),(N-L)zeros]^T$   and $\mathbf{n}_l \in N(0,N_0,\mathbf{I}_N)$ is additive noise vector. This signal is then converted to the frequency domain, which is given by: \vspace{-1.0em}
                                                                    
       \begin{equation}
       \label{fifteen}
      \mathbf{\tilde{y}}^d_l= \mathbf{F}_N \mathbf{y}^d_l
       \end{equation}
       
where $\mathbf{F}_N$ is the N-point DFT matrix.\vspace{-1.0em}
    
     \begin{equation}
       \label{sixteen}
 \begin{split}
         \mathbf{\tilde{y}}^d_l &= \sum_{j=1}^{N_e} \mathbf{H}^d_{j,l} e^d_j + \mathbf{\tilde{n}}_l \\
                       &= \sum_{j=1}^{N_e} \mathbf{H}^d_{j,l} \sum_{i=1}^{K} \mathbf{A}^i \mathbf{z}^d_{i,j} + \mathbf{\tilde{n}}_l
  \end{split}       
       \end{equation}
       
where $\mathbf{H}^d_{j,l}=diag(\mathbf{F}_N \mathbf{h}^d_{j,l})$ is the $N X N$ diagonal matrix whose diagonal elements are frequency domain coefficients between $j^{th}$ transmit antenna at eNB and $l^{th}$ UE. Let $\mathbf{\bar{A}}^i$ be the $M X N$ deallocation matrix where $\mathbf{\bar{A}}^i = (\mathbf{A}^i)^T$ . The $M$-point received signal for $l^{th}$ UE after sub-carrier deallocation is given by:\vspace{-1.0em}

\begin{equation}
       \label{seventeen}
       \begin{split}
          \mathbf{\bar{y}}^d_l &= \mathbf{\bar{A}}^i \mathbf{\tilde{y}}^d_l \\
                      &= \sum_{j=1}^{N_e} \sum_{i=1}^{K} \mathbf{\bar{A}}^i \mathbf{H}^d_{j,l} \mathbf{A}^i \mathbf{z}^d_{i,j} +\mathbf{\bar{n}_l}
       \end{split}       
       \end{equation}                                                               
                                                               
Now, the received signal for the $l^{th}$ UE on the $m^{th}$ subcarrier is given by: \vspace{-1.0em} 
                           
                             \begin{equation}
       \label{eighteen}
     \bar{y}^d_l (m) = \mathbf{H}^d_l (m) \sum_{i=1}^{K} \mathbf{P}^i_m \bar{x}^d_i (m) + \bar{n}_l (m)
       \end{equation}
       
where $\mathbf{H}^d_l (m)$  is the $1 X N_e$ frequency domain channel coefficient vector of $l^{th}$ UE on the $m^{th}$ subcarrier. $(1,j)^{th}$ entry is the $m^{th}$ diagonal element of matrix $\mathbf{H}^d_{j,l}$. $\mathbf{\bar{n}}_l (m)$  is channel noise for the $l^{th}$ UE on the $m^{th}$ subcarrier. The SVD decomposition of channel matrix  is given by \cite{sc_fdma, SVD_mimo}:
                                    
\begin{equation}
       \label{nineteen}
      \mathbf{H}^d_l (m)= U^d_{m,l} E_{m,l}^d (\mathbf{V}^d_{m,l})^H
       \end{equation}                                    
                                    
where for a single antenna UE, $U^d_{m,l}$ is a scalar such that $(\mathbf{U}^d_{m,l})^2=1$ ,$E^d_{m,l}$  is a scalar equal to $(\lambda^d_{m,l})^{1/2}$  where $\lambda^d_{m,l}$ is the eigenvalue of $\mathbf{H}^d_l(m) (\mathbf{H}^d_l(m))^H$  and $\mathbf{V}^d_{m,l}$ is a $N_e X1$  matrix containing the eigenvector corresponding to non-zero eigenvalue of $(\mathbf{H}^d_l(m))^H \mathbf{H}^d_l(m)$, which is equal to $\lambda^d_{m,l}$. 
    The received signal vector on $m^{th}$ subcarrier due to all UEs sharing the subcarriers is hence can be given by: \vspace{-0.25em}
                          \begin{equation}
       \label{twenty}
      \mathbf{\bar{y}}^d (m) = \mathbf{U}^d_m \mathbf{E}^d_m (\mathbf{V}^d_m)^H P_m \mathbf{\bar{x}}^d(m) + \mathbf{\bar{n}} (m)
       \end{equation} 
       
where $\mathbf{\bar{y}}^d (m) =[\bar{y}^d_l(m),\bar{y}^d_2(m),...,\bar{y}^d_K(m)]^T$, $\mathbf{U}^d_m= diag(U^d_{m,1},U^d_{m,2},...,U^d_{m,K})$, $\mathbf{V}^d_m = [\mathbf{V}^d_{m,1},\mathbf{V}^d_{m,2},...,\mathbf{V}^d_{m,K}]$, $\mathbf{E}^d_m=diag(E^d_{m,1}, E^d_{m,2},...,E^d_{m,K})$,$\mathbf{P}_m=[\mathbf{P}^1_m, \mathbf{P}^2_m,...,\mathbf{P}^K_m]$ ,$\mathbf{\bar{x}}^d(m)=[\bar{x}_1^d(m),\bar{x}_2^d(m),...,\bar{x}_K^d(m)]^T$and $\mathbf{\bar{n}}(m)=[\bar{n}_1(m),\bar{n}_2(m),...,\bar{n}_k(m)]^T$
 The interference from the downlink of other UEs on the $l^{th}$ UE can be completely eliminated by choosing the precoding matrix as:\vspace{-0.5em}
          
      \begin{equation}
       \label{twenty_one}
      \mathbf{P}_m=[(\mathbf{V}^d_m)^H]^+ \boldsymbol{\beta}_m
       \end{equation} 
        
 where $[(\mathbf{V}^d_m)^H]^+$ is pseudo inverse of $\mathbf{V}^d_m$, $\boldsymbol{\beta}_m = diag (\beta^1_m, \beta^2_m,..., \beta^K_m)$ defines the optimal power allocated to each of the $K$ UEs on $m^{th}$ subcarrier \cite{SVD_mimo}. The equation (\ref{twenty_three}) can be represented as: \vspace{-0.5em}

                      \begin{equation}
       \label{twenty_two}
      \mathbf{\bar{y}}^d (m) = \mathbf{U}^d_m \mathbf{E}^d_m \boldsymbol{\beta}_m \mathbf{\bar{x}}^d (m) + \mathbf{\bar{n}} (m)
       \end{equation} 
       
The received signal on $m^{th}$ subcarrier for the $l^{th}$ UE is given by:\vspace{-1.0em}
                                     
    \begin{equation}
       \label{twenty_three}
      \bar{y}^d_l (m) = U^d_{m,l} \bar{E}^d_{m,l} \bar{x}^d_l (m) + \bar{n}_l (m)
       \end{equation} 

where $\bar{E}^d_{m,l} = \bar{E}^d_{m,l} \beta^l_m$
  
In the post-processing unit, for the $l^{th}$ UE, the received signal is multiplied with $(U^d_{m,l})^H$: \vspace{-1.0em}

   \begin{equation}
       \label{twenty_four}
       \begin{split}
      \hat{y}^d_l (m) &= (U^d_{m,l})^H \bar{y}^d_l(m) \\
                      &= \bar{E}^d_{m,l} \bar{x}^d_l (m) + w_l(m)
            \end{split}
       \end{equation} 
                                               
where for a single antenna UE, $(U^d_{m,l})^H = U^d_{m,l}$  
     Using the above definition, the received signal vector for $l^{th}$ UE on the $M$ allocated subcarriers can be expressed by:
            
   \begin{equation}
       \label{twenty_five}
      \mathbf{\hat{y}}^d_l = \mathbf{\tilde{E}}^d_l \mathbf{\bar{x}}^d_l + \mathbf{w_l}
       \end{equation} 
       
where $\mathbf{\hat{y}}^d_l=[\hat{y}^d_l(1),\hat{y}^d_l(2),...,\hat{y}^d_l(M)]^T$,$\mathbf{\tilde{E}}^d_l=diag(\tilde{E}^d_{1,l},\tilde{E}^d_{2,l},...,\tilde{E}^d_{M,l}) $ , $\mathbf{\bar{x}}^d_l = [\bar{x}^d_l,\bar{x}^d_l(2),...,\bar{x}^d_l(M)]^T $ and $\mathbf{w_l}=[w_l(1),w_l(2),...,w_l(M)]^T $ . 
     This is then subjected to frequency domain MMSE equalization. The received signal vector at the output of the MMSE equalizer on the M allocated subcarriers is:  \vspace{-0.5em}
                         
   \begin{equation}
       \label{twenty_six}
      \mathbf{\hat{\bar{x}}}^d_l=((\mathbf{\tilde{E}}^d_l)^H (\mathbf{\tilde{E}}^d_l )+ N_0 I_m)^{-1} (\mathbf{\tilde{E}}^d_l)^H \mathbf{\hat{y}}^d_l
       \end{equation} 
       
This signal for the $l^{th}$ UE is then converted to time domain by an M-point IDFT operation given by:
 \vspace{-0.5em}
    \begin{equation}
       \label{twenty_seven}
     \mathbf{ \hat{x}}^d_l=\mathbf{\bar{F}}_M \mathbf{\hat{\bar{x}}}^d_l
       \end{equation} 
       
where $l=1,2,...,K$ and $\mathbf{\bar{F}}_M$  is M-point inverse IDFT matrix. This is used for decoding of the signal for the $l^{th}$  UE.    

\section{Uplink Operation}
     In Fig.\ref{ftwo}, let $\mathbf{x}^u_i$ denotes the information data block of length $M$ for $i^{th}$ UE in uplink (denoted by u):\vspace{-1.0em}
                    
   \begin{equation}
       \label{twenty_eight}
      \mathbf{x}^u_i=[x^u_i(1),x^u_i(2),...,x^u_i(M)]^T, i=1,2,...,K
       \end{equation} 
                           
The output of the M-point DFT block is given by: \vspace{-0.5em}

 \begin{equation}
       \label{twenty_nine}
      \mathbf{\bar{x}}^u_i = \mathbf{F}_M \mathbf{x}^u_i
       \end{equation} 
                   
where $\mathbf{\bar{x}}^u_i=[\bar{x}^u_i,\bar{x}^u_i(2),...,\bar{x}^u_i(M)]^T$.As discussed, $\mathbf{A}^i$ represents the $N X M$  subcarrier allocation matrix for $i^{th}$ UE, $i=1,2,…,K$. \textit{Due to channel reciprocity, the subcarrier allocation matrix for a UE in both uplink and downlink is equal}. The $N X1$  vector input to the $N$-point IDFT block for $i^{th}$ UE is given by:\vspace{-1.0em}

 \begin{equation}
       \label{thirty}
      \mathbf{d}^u_i=\mathbf{A}^i \mathbf{\bar{x}}^u_l
       \end{equation} 
        
The output of $N$-point IDFT block for the $i^{th}$ UE is given by:\vspace{-1.0em}
   
    \begin{equation}
       \label{thirty_one}
      \mathbf{s}^u_i = \mathbf{\bar{F}}_N \mathbf{d}^u_i
       \end{equation} 
       
This signal is then multiplexed into $N_r$ copies after addition of the cyclic prefix (CP). The signal, after multiplying with the complex weights,$w_{i,k}$, is transmitted in the direction of eNB w.r.t $i^{th}$ UE,.i.e., $\psi_{e,i}$, from the $N_r$ antennas with spatial response $\alpha^{k-1}_{e,i}$ for the $k^{th}$ antenna. This steers the beam in the direction of eNB according to CLMS algorithm described in the previous section. The transmitted signal from the $k^{th}$ antenna of $i^{th}$ UE is represented as follows: \vspace{-0.5em}
   
    \begin{equation}
       \label{thirty_two}
      \mathbf{s}^u_{i,k} = \alpha^{k-1}_{e,i} w_{i,k} \mathbf{s}^u_i, k=1,2,...,N_r
       \end{equation} 
       
   At the eNB (Fig.\ref{fone}), the received signal is obtained after SIC cancellation in the $N_e$ receive chains. The received signal vector of size $N X 1$  at the $j^{th}$ receive antenna due to the $K$ UEs, after removing the CP, is given by:\vspace{-1.5em}
                     
 \begin{equation}
       \label{thirty_three}
     \begin{split}
     \mathbf{y}^u_j &= \sum_{i=1}^{K} \sum_{k=1}^{N_r} \mathbf{h}^u_{j,i,k} \otimes \mathbf{s}^u_{i,k} + \mathbf{n}_{j,i} \\
           &= \sum_{i=1}^{K} \sum_{k=1}^{N_r} \mathbf{h}^u_{j,i,k} \otimes [\alpha^{k-1}_{e,i} w_{i,k} \mathbf{s}^u_i] + \mathbf{n}_{j,i}
     \end{split}
       \end{equation}                      
                     
 where $j=1,2,...,N_e$, $\mathbf{h}^u_{j,i,k}$  is the channel coefficient between $j^{th}$ antenna of eNB and $k^{th}$ antenna of $i^{th}$ UE in uplink. $\mathbf{n}_{j,i}$  is the noise introduced in the channel between $i^{th}$ UE and $j^{th}$ antenna of eNB. As we have assumed highly correlated antennas at the UEs, we take $\mathbf{h}^u_{j,i,1} \approx \mathbf{h}^u_{j,i,2}\approx...\approx \mathbf{h}^u_{j,i,N_r} \approx \mathbf{h}^u_{j,i}$. Hence, similar to the analysis in (\ref{twelve}-\ref{thirteen}) for the case of downlink and using (\ref{eleven}), the received signal in (\ref{thirty_three}) can now be represented as:\vspace{-1.0em}
    
     \begin{equation}
       \label{thirty_four}
      \mathbf{y}^u_j=\sum_{i=1}^{K}\mathbf{h}^u_{j,i} \otimes \mathbf{s}^u_i + \mathbf{n}_{j,i}
       \end{equation} 
       
where $j=1,2,...,N_e$, $\mathbf{h}^u_{j,i}=[h^u_{j,i}(0),h^u_{j,i}(1),...,h^u_{j,i}(L-1),(N-L)zeros]^T$ and $\mathbf{n}_{j,i} \in cN(0,N_0,\mathbf{I}_N)$ is the additive noise vector, which due to channel reciprocity, is equal for each pair of antenna in eNB and UE in both uplink and downlink . The output of the $j^{th}$ antenna received signal is then converted to the frequency domain by taking the $N$-point DFT, which is given by:\vspace{-1.0em}
                                                                                                      
               \begin{equation}
       \label{thirty_five}
       \begin{split}
            \mathbf{\tilde{y}}^u_j &= \mathbf{F}_N \mathbf{y}^u_j \\
                          &= \sum_{i=1}^{K} \mathbf{H}^u_{j,i} \mathbf{d}^u_i + \mathbf{\tilde{n}}_j
       \end{split}
      \end{equation} 
                                                                                               
where $j=1,2,...,N_e$ and $\mathbf{H}^u_{j,i}=diag(\mathbf{F}_N h^u_{j,i})$ is the $N X N$ diagonal matrix whose diagonal elements are the frequency domain channel coefficients between antenna of $i^{th}$ UE and $j^{th}$ receive antenna at eNB. As discussed,$\mathbf{\bar{A}}^i$ is the $M X N$ deallocation matrix where $\mathbf{\bar{A}}^i=(\mathbf{A}^i)^T$. The M-point received signal on the $j^{th}$ antenna after sub-carrier deallocation is given by:\vspace{-1.0em}
                                                                         
     \begin{equation}
       \label{thirty_six}
      \mathbf{\bar{y}}^u_j = \mathbf{\bar{A}}^i \mathbf{\tilde{y}}^u_j
       \end{equation}                                        
                                                                        
For the $m^{th}$ subcarrier, .i.e., the $m^{th}$ element of $\mathbf{\bar{y}}^u_j$, the received signal on the $j^{th}$ antenna is given by:
                                 
 \begin{equation}
       \label{thirty_seven}
      \bar{y}^u_j (m) = \sum_{i=1}^{K} H^u_{j,i}(m) \bar{x}^u_i(m) + \bar{n}_{j,i}(m)
       \end{equation}                                  
                                 
The signal received by the eNB on all the $N_e$ antennas for the $m^{th}$ subcarrier is given by:
                    
                     \begin{equation}
       \label{thirty_eight}
      \mathbf{\bar{y}}^u = \sum_{i=1}^{K} \mathbf{H}^u_i(m) \mathbf{\bar{x}}^u_i(m) + \mathbf{\bar{n}}_i(m)
       \end{equation}  
       
where $ i=1,2,...,K $, $\mathbf{\bar{y}}^u(m)=[\bar{y}^u_l(m), \bar{y}^u_2(m),...,\bar{y}^u_{N_e}(m)]^T$, $\mathbf{H}^u_i(m)=[H^u_{1,i}(m),H^u_{2,i}(m),...,H^u_{N_e,i}(M)]^T$ and $\mathbf{\bar{n}}_i(m)=[\bar{n}_{i,1},\bar{n}_{i,2},...,\bar{n}_{i,N_e}(m)]^T$  
For decoding of $l^{th}$  UE signal, the signal received given by (\ref{thirty_eight}) can be represented as:\vspace{-1.0em}
                
                 \begin{equation}
       \label{thirty_nine}
      \mathbf{\bar{y}}^u = \mathbf{H}^u_l(m) \mathbf{\bar{x}}^u_l(m) \sum_{\substack{i=1 \\ i \neq l}}^{K} \mathbf{H}^u_i(m) \mathbf{\bar{x}}^u_i(m) + \mathbf{\bar{n}}(m)
       \end{equation}  
               
where the first term represents the desired UE signal, the second term represents the co-channel interference from the other UEs and the last term $\mathbf{\bar{n}}(m)=\mathbf{\bar{n}}_l(m)+\mathbf{\bar{n}}_i(m)$  is the noise term. \par 
The received signal is subsequently passed through frequency domain MMSE equalizer. The estimated signal for the $l^{th}$ UE on the $m^{th}$ subcarrier is given by:\vspace{-1.5em}
    
      \begin{multline}
       \label{forty}
      \hat{\bar{x}}^u_l(m)=[\sigma^{-1}_{\bar{x}^u_l(m)} + (\mathbf{H}^u_l(m))^H {\mathbf{R}^u_{\bar{qq}}}^{-1}(m) \mathbf{H}^u_l(m)]^{-1} \\
       (\mathbf{H}^u_l (m))^H {\mathbf{R}^u_{\bar{qq}}}^{-1} (m) \mathbf{\bar{y}}^u(m)
             \end{multline}  
           
where $\mathbf{\bar{q}}^u(m)=\sum_{\substack{i=1 \\ i \neq l}}^K \mathbf{H}^u_l(m)\mathbf{\bar{x}}^u_i(m)+\mathbf{\bar{n}}(m)$ is the co-channel interference and noise factor for the $l^{th}$ UE. $\mathbf{R}^u_{\bar{qq}}(m)=\sum_{\substack{i=1 \\ i \neq l}}^K \mathbf{H}^u_l(m)(\mathbf{H}^u_l(m))^H \sigma^2_{\bar{x}^u_i(m)}+ N_0 \mathbf{I}_{N_e}$ is the covariance of $\mathbf{\bar{q}}^u(m)$. $\sigma^2_{\bar{x}^u_i(m)}$ is the signal power for the $l^{th}$ UE and is normalized such that  $\sigma^2_{\bar{x}^u_i(m)}=1$, hence the estimated signal term for the $l^{th}$  UE on the $m^{th}$ subcarrier, ignoring the scaling term, is given by:\vspace{-0.75em}

                                          \begin{equation}
       \label{forty_one}
      \hat{\bar{x}}^u_l(m)=(\mathbf{H}^u_l(m))^H {\mathbf{R}^u_{\bar{qq}}}^{-1} (m) \mathbf{\bar{y}}^u(m)
       \end{equation}  
       
     The estimated signal of the UEs can be estimated by the successive interference cancellation with optimal ordering (SSIC-OO) procedure defined in algorithm(\ref{euclid}) below.
     
      \begin{algorithm}
 \begin{algorithmic}[1]
\State Let $c=K$
\While{$c>1$}
\State Calculate received power for all the K UEs
\State $PW_i = ||\mathbf{H}^u_i(m)||^2$,$i=1,2,...,c$ where $\sigma^2_{\bar{x}^u_l(m)}=1$
\State Let $l = \underset{i}{\operatorname{argmax}}(PW_i)$
\State Estimate $\hat{\bar{x}}^u_l(m)$
\State $\mathbf{\bar{y}}^u(m) = \mathbf{\bar{y}}^u(m) - \mathbf{H}^u_l(m) \hat{\bar{x}}^u_l(m)$
\State $c = c-1$  
\EndWhile
\State Now let $l$ represent index for the last remaining unestimated UE
\State $\hat{\bar{x}}^u_l(m)=(H^u_l(m))^H \bar{y}^u_l(m)$, the system, ignoring the scaling term ${\mathbf{R}^u_{\bar{qq}}(m)}^{-1}=[N_0 I_{N_e}]^{-1}$, represents a single UE and multiple receive antennas at eNB with maximal ratio combining (MRC) of user symbols
\end{algorithmic}
\caption{SSIC-OO for estimating UE signal}
\label{euclid}
\end{algorithm}
 
Let the signal for the $l^{th}$  UE on all the subcarrier is given by:\vspace{-1.5em}
                                     
                                     \begin{equation}
       \label{forty_two}
      \mathbf{\hat{\bar{x}}}^u_l = [\hat{\bar{x}}^u_l(1),\hat{\bar{x}}^u_l(2),...,\hat{\bar{x}}^u_l(M)]^T
       \end{equation}  
       
This signal for the $l^{th}$ UE is then converted to time domain by an $M$-point IDFT operation given by:\vspace{-0.5em}
    \begin{equation}
       \label{forty_three}
 \mathbf{\hat{x}}^u_l=\mathbf{\bar{F}}_M \mathbf{\hat{\bar{x}}}^u_l, l=1,2,...,K
       \end{equation}  
       
This is used for decoding of the signal for the $l^{th}$  UE. 

\section{Simulation Results}
The advantage of using SC-FDMA for downlink instead of OFDMA in terms of BER performance is analyzed in \cite{sc_fdma}. To validate the inclusion of SIC design for FD system in our architecture, we have carried out MATLAB simulations for BER performance in downlink and uplink. For simulation, we have considered an FD eNB with four antennas ($N_e=4$) and two FD UEs ($K=2$) with four antennas each ($N_r =4 $) and $PW_2 > PW_1$. The DoA of eNB w.r.t two stationary UEs, i.e, UE1 and UE2 is $10^o$ and $60^o$ respectively \footnote{Such that the UEs are at the null of each others beampattern.}. The issues due to mobility of UEs are considered in the next part of the work. For smart antenna beamforming at UEs, the CLMS algorithm with root music algorithm is used. The Ghorbani Model and thermal noise (Noise temperature = $290K$) is used for modeling the non-linearity which is introduced to the complex baseband SC-OFDM symbols. The channel between each antenna of eNB and UEs’ is taken as frequency selective with $L=10$ and uniform power delay profile (UPDP). The modulation scheme used is 16-QAM (no coding). The bandwidth allocated to the UEs is taken to be 3MHz which is split into 256 subcarriers, out of which 180 subcarriers are occupied by the UEs. A cyclic prefix of duration 4.69μs is used. \textit{The UEs operate in downlink and uplink on all the subcarriers}. \par
\begin{figure}
\centering
\includegraphics[width=2.70in ,height=1.65in]{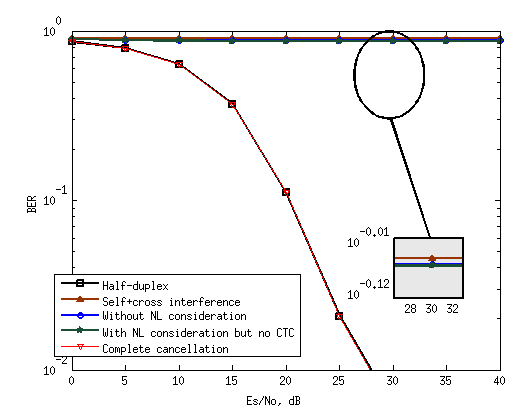}
\vspace{-0.5em}
   \caption{BER performance at UE1 for FD downlink }
   \label{ffour}
   \vspace{-1.0em}
\end{figure}

   For downlink, the performance at UE1 is considered for received complex SC-FDMA symbols from eNB. In Fig.\ref{ffour}, the effect of SIC on the receiver performance at UE1 is shown. It is observed that without SIC the receiver has zero throughput. A similar result is obtained when SIC is attempted without taking into consideration the non-linearity (NL) components. No improvement in receiver performance is observed by including the NL distortion components in SIC without the cross-talk cancellation (CTC). The receiver performance is equivalent to half-duplex (HD) performance when we considered both self and cross talk along with NL distortion components for SIC.  \par
\begin{figure}
\centering
\includegraphics[width=2.75in ,height=1.65in]{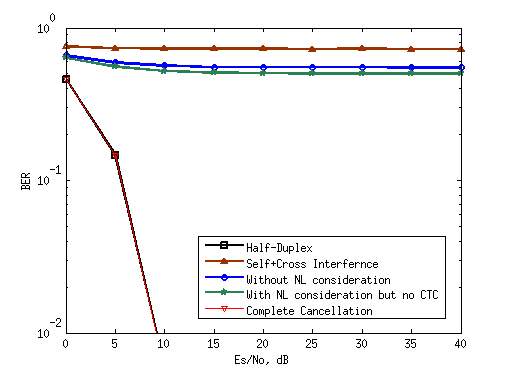}
\vspace{-0.5em}
   \caption{BER performance at eNB for FD uplink}
   \label{ffive}
   \vspace{-1.0em}
\end{figure}

   
    For uplink, the performance at multiple antenna eNB is considered for received complex SC-FDMA symbols from UE1. In Fig.\ref{ffive}, the effect of SIC on the receiver performance of the eNB is shown. The SIC analysis is similar to downlink, but there is an additional diversity gain introduced due to multiple antennas in eNB according to the algorithm(\ref{euclid}). Comparing the BER performance at UE (Fig.\ref{ffour}) and eNB (Fig.\ref{ffive}), it is observed that due to the additional diversity gain, there has been a nearly 18dB gain for the eNB over the BER performance of UE at BER $10^{-2}$. In case of uplink, the effect of CCI has negligible impact as eNB employs SSIC-OO to cancel out interference of signal of one UE on the signal of other UEs. Similar analysis can be done by taking UE2 into consideration. \par
\begin{figure}
\centering
\includegraphics[width=2.75in ,height=1.50in]{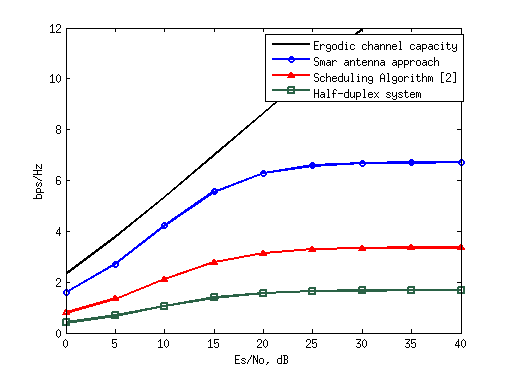}
\vspace{-0.5em}
   \caption{Overall Spectrum efficiency per cell in downlink for various scheduling approaches}
   \vspace{-0.45em}
   \label{fsix}
   \vspace{-1.5em}
\end{figure}

     In terms of the overall spectral efficiency (SE) per cell , the proposed FD eNB and FD UE transceiver along with the smart antenna technique, helps in achieving higher performance\footnote{The smart antenna technique helps more UEs to be scheduled simultaneously using same spectrum resource in downlink and uplink.} as compared to HD time division duplexing (TDD) system and scheduling algorithm proposed in \cite{FD_small}. The SE per cell in downlink for two UEs, assuming total SIC, is shown in fig.\ref{fsix}. For the HD TDD system, only one UE is scheduled for uplink and downlink in alternate time slots. For scheduling algorithm \cite{FD_small}, two HD UEs (with FD eNB) is considered such that, for a time slot, these are scheduled simultaneously in reverse directions (one in uplink and other in downlik) with their direction interchanging in every consecutive time slot \footnote{ Due to limitation in space the result is not discussed in detail here and will be included in the future work along with analysis for uplink SE.}.    \par
    
\section{Conclusion}
 In this paper, we proposed design for FD multiple antenna eNB and FD UE transceiver units and discuss the corresponding uplink and downlink operations. The smart antenna technique is used to avoid interference at downlink of a UE from the uplink signal of other UEs operating in the same subcarriers. Finally, simulations for uplink and downlink operations along with overall SE per cell in downlink for the proposed FD system is analyzed. In the future work, we will try to relax assumptions made in the paper and optimize the designs in term of energy efficiency. Further, we will exploit utility of mmWaves in small cell FD deployment.





%



\end{document}